\documentclass{iaus}
\usepackage{graphicx}

\title[MilkyWay@home] 
{MilkyWay@home: Harnessing volunteer computers to constrain dark matter in the Milky Way}

\author[Newberg, Newby, Desell, Magdon-Ismail, Szymanski \& Varela]   
{Heidi Jo Newberg,$^1$
Matthew Newby,$^1$
Travis Desell,$^2$
Malik Magdon-Ismail,$^3$
Boleslaw Szymanski,$^3$
\and Carlos Varela$^3$}

\affiliation{$^1$Dept. of Physics, Applied Physics, and Astronomy, Rensselaer Polytechnic Institute, \\ 110 8th St., Troy, NY 12180, \\ email: {\tt heidi@rpi.edu} \\[\affilskip]
$^2$Dept. of Computer Science, U. of North Dakota, \\ Grand Forks, ND 52802 \\[\affilskip]
$^3$Dept. of Computer Science, Rensselaer Polytechnic Institute, 
\\ 110 8th St., Troy, NY, 12180}

\pubyear{2013}
\volume{xxx}  
\pagerange{119--126}
\jname{Title of your IAU Symposium}
\editors{A.C. Editor, B.D. Editor \& C.E. Editor, eds.}
\begin{document}

\maketitle

\begin{abstract}
MilkyWay@home is a volunteer computing project that allows people from every country in the world to volunteer their otherwise idle processors to Milky Way research.  Currently, more than 25,000 people (150,000 since November 9, 2007) contribute about half a PetaFLOPS of computing power to our project.  We currently run two types of applications: one application fits the spatial density profile of tidal streams using statistical photometric parallax, and the other application finds the $N$-body simulation parameters that produce tidal streams that best match the measured density profile of known tidal streams.  The stream fitting application is well developed and is producing published results.  The Sagittarius dwarf leading tidal tail has been fit, and the algorithm is currently running on the trailing tidal tail and bifurcated pieces.  We will soon have a self-consistent model for the density of the smooth component of the stellar halo and the largest tidal streams.  The $N$-body application has been implemented for fitting dwarf galaxy progenitor properties only, and is in the testing stages.  We use an Earth-Mover Distance method to measure goodness-of-fit for density of stars along the tidal stream.  We will add additional spatial dimensions as well as kinematic measures in a piecemeal fashion, with the eventual goal of fitting the orbit and parameters of the Milky Way potential (and thus the density distribution of dark matter) using multiple tidal streams.
\end{abstract}

\firstsection 
\section{Fitting the spatial density of tidal streams with statistical photometric parallax}

It is only in this century that we have known that the Milky Way's stellar halo has spatial substructure, due to the tidal disruption of dwarf galaxies and globular clusters \cite[(Newberg et al. 2002; Belokurov et al. 2006)]{nyetal02,belokurov}.  This discovery was made possible due to the availability of data from high quality, large area surveys of the sky - primarily the Sloan Digital Sky Survey (SDSS; York et al. 2000), but also the Two Micron All Sky Survey (2MASS; Majewski et al. 2003).  The spatial substructure was originally found ``by eye" in maps of star density, which are constructed from the angular sky positions of the stars in the tracer population, and an estimate of their distance.  Streams continue to be found ``by eye" even today, though more sophisticated techniques have been developed for pulling out fainter structures by eye \cite[(Grillmair 2009)]{grillmair4streams}, and more often streams are first detected from velocity substructure \cite[(Martin et al. 2013)]{PSS}.

Although we have not yet demonstrated successful methods for finding tidal streams in an automated fashion, we have made progress in describing the density structure of individual streams, and in particular the tidal streams from the Sagittarius dwarf galaxy.  This has been a difficult task for two reasons: (1) the tidal streams do not have simple geometric shapes, and (2) the highest density tracer populations we have observed are F turnoff stars, which have a large range of absolute magnitudes and are therefore not good indicators of distance.

The second of these problems has been addressed by the technique of statistical photometric parallax \cite[(Newberg 2013)]{statphot}, in which statistical knowledge of the absolute magnitudes of stellar populations is used to determine the underlying density distributions that these stars trace.  Although the absolute magnitude of the turnoff in general depends on the age and metallicity a stellar population, it turns out that the age-metallicity relationship in the Milky Way leads to a nearly constant absolute magnitude distribution for spheroid turnoff stars \cite[(Newby et al. 2011)]{newby11}; although older populations should have fainter turnoffs, they are also generally more metal-poor, which pushes the turnoff brighter by a nearly equal amount.  In addition, there is a fairly uniform distribution of absolute magnitudes in a color-selected sample (assuming negligible color errors) of turnoff stars from the full range of stellar populations in halo globular clusters.  

To determine the likelihood of a particular model stellar density, then, one follows the following procedure: (1) For each stellar component, one assumes a parameterized model.  (2) The spatial density expected from the model is transformed to $(l,b,g)$ coordinates, where g is the apparent magnitude, assuming all of the stars have the same absolute magnitude.  (3) The density is then convolved with the absolute magnitude distribution of the tracer population, which smears the model out along the line-of-sight.  (4) The expected distribution is multiplied by the fraction of stars that are expected to be observed, which is usually a function of apparent magnitude (a lower fraction of stars are detected near the survey limit).  (5) The resulting distribution is normalized so that the integrated probability of finding a star in the entire observed volume is one.  (6) The final probability distribution function (PDF) is the sum of the fraction of stars in each component (these are also model parameters) times the normalized distribution, summed over the number of components in the model.  (7) The likelihood of this particular parameterized model is the product of the probability density function evaluations at the location of each observed tracer star in the survey.  One must use an optimization technique to find the parameters of the model that best fit the data.

One year after the lumpy nature of the stellar halo was discovered (see \cite[Newberg et al. 2002]{ntetal02}, Figure 1), Newberg, Magdon-Ismail and a graduate student attempted to use statistical photometric parallax to determine the density structure of the debris in that dataset.  We devised a density model for tidal streams that fit the position, width and density of the stream in $2.5^\circ$-wide segments, matching the width of the SDSS data stripes.  This allowed the properties of the stream to vary along its length, and the position and distance of the center of the tidal debris stream to vary arbitrarily across the sky.  We eventually determined that using conjugate gradient descent to optimize the likelihood on a single processor would take more than a year.  The difficulty was the length of time required to compute the integral of the PDF over the volume observed.

The solution was the parallelization of the algorithm.  Then graduate student Travis Desell converted the code to run in parallel on a computer cluster, and on Rensselaer's IBM BlueGene supercomputer, and on the Berkely Open Infrastructure for Network Computing \cite[(BOINC, Anderson 2004)]{boinc}.  The algorithm and the proof of concept was published in \cite[Cole et al. (2008)]{cole08}.  The density of the Sagittarius (Sgr) dwarf leading tidal tail was published in \cite[Newby et al. (2013)]{newby13}.  We did not succeed in automated detection of new tidal streams.  However, the maximum likelihood algorithm has given us information about the width of the stream as a function of position along the stream, and a more robust way of calculating the density of stars along the stream, especially in cases where some of the stream stars are missed due to the limiting magnitude of the survey.  More importantly, we have a method for separating stars with the density structure of the tidal stream from the smooth component of the stellar halo.

We are currently fitting the Sgr dwarf trailing tidal tail, the bifircated pieces of both the leading and trailing tidal tails, and the Virgo overdensity.  We will then fit a smooth model to the remaining spheroid stars that are not in major tidal streams, to genenerate a self-consistent denisty model for the halo.

\section{MilkyWay@home Infrastructure}

\begin{figure}[b]
\begin{center}
 \includegraphics[width=5.25in]{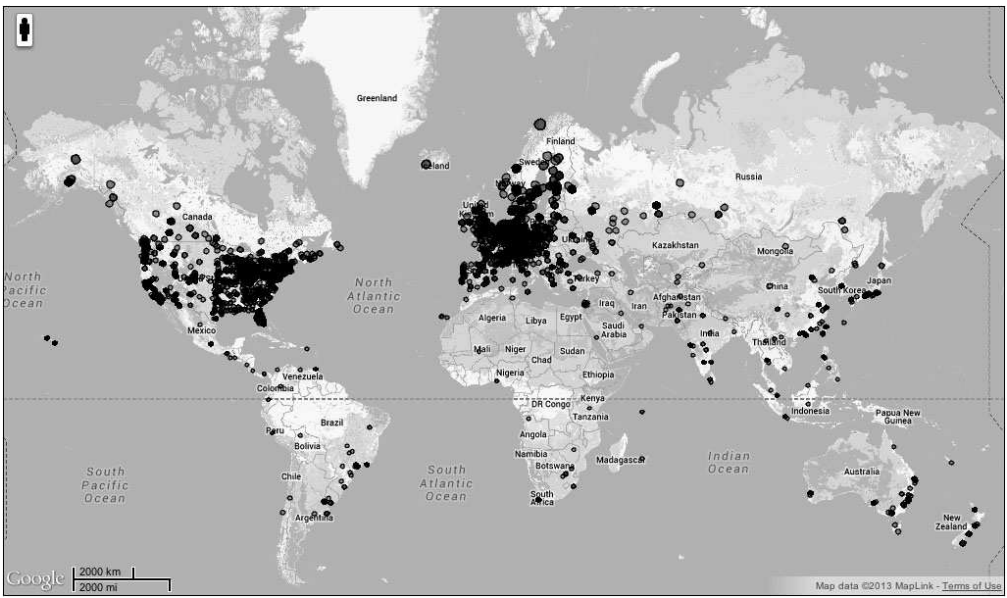} 
 \caption{Worldwide locations of processors donated to MilkyWay@home.  This image was generated on June 19th, 2013, using a sample of 10,212 active volunteered computing hosts from MilkyWay@Home, each host having returned valid results within approximately one week.  The locations of these hosts' external IP addresses was determined using a GeoIP database (http://www.hostip.info/), and then plotted on the globe using Google maps (http://maps.google.com/).}
 \label{fig1}
\end{center}
\end{figure}

Our most exciting parallel platform is MilkyWay@home.  We began test operations on November 9, 2007.  Although most of our volunteers only supply computing power, many are part of a community that follows the scientific progress of a project, participate in a variety of on-line forums, and even help with solving technical and coding issues.  A few have donated small amounts of money and hardware.  It was a volunteer who first ported our code to GPUs, and showed us the enormous power of these devices.  He also caused an outcry on our public forums, since he was able to amass BOINC ``credits" at a much faster rate than anyone else.  Apparently, the amassing of BOINC credits (which are only posted on the web and cannot be used in any way), is a serious issue.  We also have a volunteer moderator who polices the forums to ensure no one is posting inappropriate material, and informs us when the server is malfunctioning.  Since inception, 162,382 people from 206 countries (193 members of the United Nations) world-wide have contributed their otherwise unused computer cycles to our project.  At any given time, $\sim 25,000$ people from $150$ countries are crunching our work units (Fig.~\ref{fig1}).

MilkyWay@home is a tremendous computing resource.  We operate a server that generates parameter sets we wish to evaluate, sends one parameter set as a ``work unit" to each available volunteer, then receives the result (the likelihood) back from each volunteer when the computation is complete.  The result is stored in a database, and another work unit (with a new set of parameters to try) is sent out.  It currently delivers about half a PetaFLOPS of computing power, down from a peak of 2 PetaFLOPS shortly after GPU code was released.  To put this in perspective, MilkyWay@home had the processing power equivalent to the 45$^{\rm th}$ fastest supercomputer in the world in November 2012.

This computing power comes at a hefty price, however.  Our server generates work units (in this case likelihood calculations for one set of model parameters) that are sent out to a very heterogeneous set of processors.  We compile our code for sixteen platforms, and respond to questions and problems from our volunteers when there are bugs.  It takes us months to release and re-release new algorithms before the bugs are solved on all platforms.  In addition to algorithm enhancements, our code must be updated in conjunction with any updates in the BOINC infrastructure.  We require fault-tolerant optimization algorithms that work in a highly asynchronous, heterogeneous computing environment.  That is, sometimes (either through hardware/software malfunction or malicious intent) the results sent back from a volunteer are incorrect.  Also, the time to finish a work unit is highly variable since it depends on the hardware platform that is doing the computation and on the availability of the hardware (which could be turned off for the weekend or working on other tasks).  We need to check the validity of a fraction of the work units (by sending them to multiple volunteers) and track the veracity of the results from each volunteer, so that we reject results from consistently incorrect platforms.

The advantages of MilkyWay@home are that we have our very own supercomputer and worldwide outreach program.  The only hardware we need to upgrade on a regular basis is a server and a development platform; the volunteers upgrade their hardware at their own expense.  We have shown \cite[(Newby et al. 2013)]{newby13} that the results we get from MilkyWay@home are as good, and usually better, than the results we got from one rack of a supercomputer using conjugate gradient descent.  The time to converge to the correct solution is about the same for these two platforms, but we can run many more simultaneous jobs on MilkyWay@home (and we don't have to wait for time in the queue).

\section{Comparing $N$-body simulations to the spatial density of tidal streams}

The availability of our own supercomputer has made it possible for us to make bigger plans.  We would like to be able to use the tidal debris streams to optimize parameters in the mass distribution of the Milky Way, and learn about the orbital parameters of the dwarf galaxies and globular clusters that are the progenitors of the tidal debris that we see today.  Most of the previous work fitting parameters of the Milky Way potential focusses on the Sagittarius dwarf tidal stream \cite[(for example see Law \& Majewski 2010)]{lm2010}.  \cite[Koposov, Rix \& Hogg (2010)]{koposov} fit orbits to the stars in the GD-1 stream .  \cite[Willett (2010)]{willettthesis} attempted to fit halo parameters with simultaneous orbit fits to three tidal streams.  Recent work by \cite[Binney (2008) and Sanders \& Binney (2013)]{binney, sanders} points out that tidal streams of dwarf galaxies do not follow the orbits of the dwarf galaxy, and they work towards new methods for measuring halo parameters from tidal streams.  We believe that with MilkyWay@home we will eventually be able to fit multiple tidal streams to $N$-body simulations to simultaneously constrain the properites of the dwarf galaxies and their orbits, and the distribution of Milky Way dark matter.

We have a test $N$-body simulation running on MilkyWay@home, that uses a version of the \cite[Barnes and Hut (1986)]{barnshut} tree code.  We currently fit only the properties of the progenitor, which is modeled as a \cite[Plummer (1911)]{plummer} sphere with 100,000 bodies.  We fit the parameters of both dark matter and stars in the progenitor, and we also fit the simulation time assuming a fixed Milky Way potential, and with a fixed progenitor orbit.  The bodies corresponding to stars are sub-selected from the bodies in the original Plummer sphere.  We currently fit only the density of stars along the stream (comparing only the stars in the simulation with the observed stars in the stream), using an Earth-Mover Distance method; the similarity of two normalized histograms is measured by the number of items that must be moved and the distances they must move.  We also include a cost function for having different numbers of stars in the observed and simulated histograms.  In the future we plan to fit the density of the stream in at least two dimensions, and also the radial velocities and velocity dispersions of the stream stars.  Recent studies of $N$-body simulations fit by hand to the Cetus Polar Stream \cite[(Yam et al. 2013)]{yamCPS} suggest that the dwarf galaxy properties might be better fit if the width or velocity dispersion of the stream is fit in conjunction with the stellar density along the stream.

\section{Future Plans}

We are developing MilkyWay@home to constrain the potential of the Milky Way galaxy using tidal streams.  The application that fits the density distribution of tidal debris is well developed, but we are in the process of implementing an improved algorithm.  While the absolute magnitude distributions of F turnoff stars are not expected to change with distance in the Milky Way halo, the stellar population that is sampled in a color-selected sample changes dramatically near the survey limit as the color errors become large \cite[(Newby et al. 2011)]{newby11}.  We have recently released an algorithm that includes this effect in the likelihood calculation.  This will allow us to create a more accurate description of the spatial density of each stellar substructure in the halo, so that the densities sum to the actual observed spatial density of halo stars.

We can then use the measured spatial densities of the stars in the tidal streams, along with data on the kinematics of the stream stars, to constrain the gravitational potential (and thus the spatial density of dark matter) of the Milky Way and the properties of the progenitor dwarf galaxies (including their dark matter content).  We do this by varying the parameters in the $N$-body simulations of the tidal disruption of the dwarf galaxy progenitor until we generate a stream with the correct spatial and kinematic properties.  Initially, the only kinematic information we will have is sparsely sampled line-of-sight velocities.  Once data from Gaia is available, we will be able to fit proper motions as well.

Fitting models with a large number of parameters to data, such as is described in this proceedings, will be of growing importance as we increase the amount of data that is available, for example from surveys such as LAMOST and Gaia.  With small amounts of data, the spatial density of the halo seems to be well fit by a three parameter power law.  With millions of stars, many complex substructures are observed; and the stellar halo is shown to be a very poor fit to a simple power law.

\begin{discussion}

\discuss{P. Bonifacio}{How did you advertise your project and recruit volunteers?  The number of active computers you have is very impressive!}

\discuss{Newberg}{Our server is part of the Berkeley Open Infrastructure for Network Computing (BOINC).  When we put a new server in their system, anyone who is part of that is given the opportunity to select MilkyWay@home as a donor for their compute cycles.  We did not need to advertise MilkyWay@home ourselves.}

\discuss{Robyn Sanderson}{Do you plan to extend stream searches to multiple dimensions as new data (Gaia) arrives?}

\discuss{Newberg}{We have given up on searching for tidal streams with this system, though we are fitting major streams that are known.  We do plan to add radial velocity constraints to the application that matches known streams to $N$-body simulations, and as better proper motions become available (from Gaia) we will also consider adding these constraints.} 

\discuss{Anonymous}{Do the volunteers contribute more than their computer cycles to the project?}

\discuss{Newberg}{The volunteers do not currently contribute directly to the science.  We have submitted proposals to fund the addition of a mechanism for volunteers to see the likelihood space as a function of the parameters we are fitting, so they can select the parameters they would like to try (rather than using the ones we generate from our algorithm), but these proposals so far have not been funded.  We also have plans to tell the volunteers if their computer has generated more likely parameters than any other so far.  Our volunteers have contributed to our code development - in fact it was a volunteer that wrote the first GPU version of our software and showed us how much faster it runs - and to fixing bugs on the many platforms that our software currently runs on.  They also communicate with us on the forums about the science that is being done and a few of them have donated small amounts of money or GPUs.}

\end{discussion}


\begin{thebibliography}{}

\bibitem[Anderson(2004)]{boinc} Anderson, D.~P.\ 2004, in: (Rajkumar Buyya, ed.), \textit{Fifth IEEE/ACM International Workshop on Grid Computing}, Proc. IEEE Computing Society, p.\ 4

\bibitem[Barnes \& Hut(1986)]{1986Natur.324..446B} Barnes, J., \& Hut, P.\ 1986, \textit{Nature}, 324, 446 

\bibitem[Belokurov et al.(2006)]{belokurov} Belokurov, V., Zucker, D.~B., Evans, N.~W., et al.\ 2006, \textit{ApJ} (Letters), 642, L137 

\bibitem[Cole et al.(2008)]{cole08} Cole, N., Newberg, H.~J., Magdon-Ismail, M., et al.\ 2008, \textit{ApJ}, 683, 750 

\bibitem[Grillmair(2009)]{grillmair4streams} Grillmair, C.~J.\ 2009, \textit{ApJ}, 693, 1118 

\bibitem[Koposov et al.(2010)]{koposov} Koposov, S.~E., Rix, H.-W., \& Hogg, D.~W.\ 2010, \textit{ApJ}, 712, 260 

\bibitem[Law \& Majewski(2010)]{lm2010} Law, D.~R., \& Majewski, S.~R.\ 2010, \textit{ApJ}, 714, 229 

\bibitem[Majewski et al.(2003)]{2003ApJ...599.1082M} Majewski, S.~R., Skrutskie, M.~F., Weinberg, M.~D., 
\& Ostheimer, J.~C.\ 2003, \textit{ApJ}, 599, 1082

\bibitem[Martin et al.(2013)]{PSS} Martin, C., Carlin, 
J.~L., Newberg, H.~J., \& Grillmair, C.\ 2013, \textit{ApJ} (Letters), 765, L39 

\bibitem[Newberg et al.(2002)]{nyetal02} Newberg, H.~J., Yanny, 
B., Rockosi, C., et al.\ 2002, \textit{ApJ}, 569, 245 

\bibitem[Newberg(2013)]{2013IAUS..289...74N} Newberg, H.~J.\ 2013, in: (Richard de Gris ed.), \textit{Advancing the Physics of Cosmic Distances}, Proc. IAU Symposium No. 289 (Cambridge University Press), p.\ 74 

\bibitem[Newby et al.(2011)]{newby11} Newby, M., Newberg, H.~J., Simones, J., Cole, N., \& Monaco, M.\ 2011, \textit{ApJ}, 743, 187

\bibitem[Newby et al.(2013)]{newby13} Newby, M., Cole, N., Newberg, H.~J., et al.\ 2013, \textit{AJ}, 145, 163 

\bibitem[Plummer(1911)]{1911MNRAS..71..460P} Plummer, H.~C.\ 1911, \textit{MNRAS}, 71, 460 

\bibitem[Willett(2010)]{willettthesis} Willett, B.~A.\ 2010, Ph.D.~Thesis

\bibitem[Yam et al.(2013)]{yamCPS} Yam, W., Carlin, J.~L., Newberg, H.~J., et al.\ 2013, \textit{ApJ}, submitted

\bibitem[York et al.(2000)]{2000AJ....120.1579Y} York, D.~G., Adelman, J., Anderson, J.~E., Jr., et al.\ 2000, \textit{AJ}, 120, 1579

\end{thebibliography}
\end{document}